\documentclass[preprint]{aastex}
\usepackage{emulateapj5}

\begin{document}
\title{Lower Limits on Lorentz Factors in Gamma Ray Bursts}
\author{Yoram Lithwick and Re'em Sari}
\affil{130-33 Caltech, Pasadena, CA 91125; yoram@tapir.caltech.edu,
sari@tapir.caltech.edu}
\begin{abstract}
As is well-known, the requirement that gamma ray bursts
be optically thin to high energy photons yields
a lower limit on the Lorentz factor ($\gamma$) of the
expansion.  In this paper, we provide a simple derivation
of the lower limit on $\gamma$ due to the annihilation
of photon pairs, and correct the errors in some of the
previous calculations of this limit. 
We also derive a second limit on $\gamma$
due to scattering of photons by pair-created electrons
and positrons.  For some bursts, this limit is the more 
stringent. 
In addition, we show that a third limit on $\gamma$, which is obtained
by considering the scattering of photons by electrons
which accompany baryons, is nearly always less important
than the second limit.
Finally, we evaluate these limits for a number of bursts.
\end{abstract}
\section {INTRODUCTION}

Many gamma ray bursts (GRB's) emit photons with very
high observed energies ($\gg$ 1 MeV). 
If the  expansions of the bursts were non-relativistic, 
then the optical depth to the high energy  photons
would be so large that these photons could not be observed.
This is the  ``compactness problem'' (e.g. \cite{pir99}).

Three basic processes contribute to the optical 
depth of high energy photons: (A) annihilation of pairs of 
photons into  $e^{\pm}$ pairs; 
(B) scattering of photons by either the
$e^+$ or the $e^-$ created by the annihilation of 
photon pairs;
(C) scattering of photons
by the electrons  which are associated with baryons that 
may be present in the exploding ejecta. 

Since the optical depth from each of these three processes
decreases with increasing Lorentz factor ($\gamma$) of
the expansion,  the requirement that the burst be optically
thin yields a lower limit on $\gamma$.
In the following sections, we calculate the
appropriate lower limits on $\gamma$.
We then evaluate these lower limits for several bursts during which
high energy photons were observed. 

\section{ LIMIT A: FROM PHOTON ANNIHILATION}\label{sec:method1}

It is assumed that the photon spectrum
at high energies is a truncated power law.  
This can be regarded as an approximation to the
high energy portion of the popular 
\cite{bmf+93} parameterization.  Thus, the spectrum 
(i.e. the observed number of photons per unit time per unit area per unit
energy) 
is given by 
$fe^{-\alpha}  , \  e_{\rm min}<e<e_{\rm max}$. 
Here $e$ is the observed photon energy, and,
unless explicitly stated otherwise, all energies throughout
this paper refer to their values in the observer's reference
frame.
The factor $f$ is a normalization factor.
The exponent $\alpha$ is typically between 2 and 3; 
a value greater than 2
implies that most of the energy is in low-energy photons.
The energy $e_{\rm max}$ is the highest energy for which photons
were observed. Very often,
it is the  energy
above which the photon flux is too small to be detected 
\citep{cds98}.
The spectrum  turns over below $e_{\rm min}$, which is typically
around 0.1-1 MeV, although the exact value will not be needed 
in this paper.

Of particular interest in the calculations below will be the total
number of emitted photons which have
energies greater than $e$, if the
burst emits for an observed time $\delta T$
(where $\delta T$ is set by the variability timescale as seen 
in BATSE light curves, for example).
This number, $N_{>e}$,  
is obtained by integrating the spectrum
with respect to $e$, and then multiplying
by $\delta T$ and $4\pi d^2$ (where $d$ is the distance to the GRB),
yielding
\begin{equation}
N_{>e}=4\pi d^2 f\delta T e^{-\alpha+1}/(\alpha-1) \ .
\label{eq:ngte}
\end{equation}
Throughout this paper, we assume spherical symmetry (at least
within the beaming angle $1/\gamma$).
In addition, for clarity, we neglect corrections due to cosmological redshift
in the text.  
However, these
corrections will be included in table~(1) 
which summarizes the main results.

Now, in the frame of the emitting material, where the photons
are assumed to be roughly isotropic, a photon with energy
$e'$ can annihilate a second photon with energy larger than 
$(m_e c^2)^2/e'$, yielding an electron-positron pair, 
where $m_e$ is the electron mass.
When the energy of the second photon is
around $(m_e c^2)^2/e'$, then
the cross-section for this process is approximately the 
Thomson cross-section, $\sigma_T$. The cross-section
falls off as a power-law of the annihilating photon energy 
if its energy is significantly
larger than this value, and it is zero when its energy is below 
this value. The combination of these effects leads to an
averaged cross section of about $\frac{11}{180}\sigma_T$, assuming an 
$\alpha=2$ spectrum \citep{sve87}.

If the emitting material is moving towards the observer
with a Lorentz factor $\gamma$, then the photons 
are blueshifted
by $\gamma$. Thus, a photon with detected
energy $e=\gamma e'$ can only annihilate photons whose detected
energies are larger than 
$(\gamma m_e c^2)^2/e$. 
Since most of the photons are at small energies, the photon
with the highest energy will be most susceptible to annihilation
by other photons.
Thus, a lower limit on the Lorentz factor
can be obtained by requiring that the photon with energy $e_{\rm max}$
will have optical depth smaller than unity.

To calculate this optical depth, we note that when
the expansion is spherically symmetric (at least within the beaming
angle $1/\gamma$), 
the burst expands to a radial distance
of about $\gamma^2 c \delta T$ \citep{pir99}.
This is true (within numerical factors of order unity) independent
of the details of either the expansion scenario or of 
whether the emission is
from internal or external shocks.

Now, if $e_{\rm max,an}$
is defined as the energy of the photon which annihilates $e_{\rm max}$,
i.e. $e_{\rm max,an}\equiv (\gamma m_e c^2)^2/e_{\rm max}$, then there
are $N_{>e_{\rm max,an}}$ photons which can annihilate the $e_{\rm max}$
photon.
Thus, the optical depth is 
\begin{equation}
\tau={{11\over 180}\sigma_TN_{>e_{\rm max,an}}
\over 4\pi(\gamma^2c\delta T)^2 }\ .
\label{eq:tau}
\end{equation}

Finally,  inserting the expression 
for $N_{>e}$ (equation~\ref{eq:ngte}) into the above equation yields
\begin{equation}
\tau=
\hat{\tau}
\Big(
{e_{\rm max}\over m_e c^2}
\Big)^{\alpha-1}
\gamma^{-2\alpha-2} \ ,
\label{eq:tau1}
\end{equation}
where
\begin{equation}
\hat{\tau}\equiv
{{11\over 180}\sigma_T d^2  (m_e c^2)^{-\alpha+1}f\over c^2 \delta T (\alpha-1)}
\label{eq:tauhat}
\end{equation} 
is a dimensionless quantity which will appear again below.
Physically, it is the optical depth for a photon with
energy $m_ec^2$ in a mildly relativistic fireball
(i.e. with expansion speed $v\approx c$,  but with
$\gamma$ not much larger than unity). 
It is evaluated with observationally plausible values
in equation~(\ref{eq:numtauhat}). 

The requirement that $\tau<1$ leads to the limit
\begin{equation}
\gamma>\hat{\tau}^{1\over 2\alpha+2}\Big(
{e_{\rm max}\over m_ec^2}\Big) ^{\alpha-1\over 2\alpha+2} \ .
\label{eq:gamma1}
\end{equation}

The interpretation of the above calculation can be facilitated
by defining a characteristic photon energy, $e_{\rm thick}$.
This energy is defined as follows:
the number of photons  with energy greater than $e_{\rm thick}$ is
such that, if each of these photons were assigned a cross-section
equal to $11/180$ of the Thomson
cross-section, 
then they would be optically thick, i.e., 
\begin{equation}
1\equiv{
{11\over 180} \sigma_TN_{>e_{\rm thick}}
\over
4\pi(\gamma^2c\delta T)^2} \ , \ {\rm so\  that} \ \ 
{e_{\rm thick}\over m_e c^2}=\hat{\tau}^{1\over \alpha-1}
\gamma^{-{4\over \alpha-1}} \ .
\label{eq:ethick}
\end{equation}
Since the number of photons decreases with increasing energy,
the requirement that the optical depth of the photon with
energy $e_{\rm max}$ be less than unity is
simply the requirement that $e_{\rm max,an}>e_{\rm thick}$---from
which equation~(\ref{eq:gamma1}) follows directly.

Table (1) summarizes the main results of this paper.
In this table, the corrections due to cosmological redshift ($z$) have
been added. These corrections may be obtained by 
considering equation~(\ref{eq:tau}).  There
are three redshift effects.
(i) In calculating $N_{>e_{\rm max,an}}$, one needs to convert
from the number of observed photons per unit
area to the total number of photons at the source: instead of multiplying
by $4\pi d^2$, one should multiply by the appropriate surface 
area of a sphere which is centered
at the burst, i.e. by
$4\pi [d_L/(1+z)]^2$, where
$d_L$ is the luminosity distance.
(ii) Due to cosmological time dilation, $\delta T\rightarrow \delta T/(1+z)$
in the denominator of equation~(\ref{eq:tau}). Note that when (i) and (ii)
are combined, the explicit factors of $1+z$ cancel, although there is 
still a redshift dependence through the luminosity distance. 
(iii) Due to the redshift of photon energies, the photon which annihilates
the $e_{\rm max}$ photon need only have an energy greater than 
$((\gamma/(1+z) m_e c^2)^2/e_{\rm max}$ in order to exceed
the threshold for pair creation, i.e. 
$e_{\rm max,an}\rightarrow e_{\rm max,an}/(1+z)^2$.
When this is inserted this into $N_{>e_{\rm max,an}}$ in
equation~(\ref{eq:tau}), and combined with (i) and (ii), the result is
that $\tau\propto d_L^2(1+z)^{2\alpha-2}$.

\subsection{Comparison With Other Work}

The physical mechanism presented in this section has been used
in a number of previous papers to place a lower limit
on the Lorentz factor.  However, in most of these
papers, we found that the dependence 
of the optical depth on the Lorentz factor was incorrect.
This incorrect dependence results in a limit on $\gamma$
which is too large by a factor of, typically, 2 or 3.
The correct dependence is given in equation~(\ref{eq:tau1}),
i.e., $\tau\propto\gamma^{-2\alpha-2}$.

Both \cite{feh93} and \cite{wl95} considered the
case when $\alpha=2$, and both obtained
$\tau\propto\gamma^{-4}$ instead of the above scaling 
$\tau\propto\gamma^{-6}$ when $\alpha=2$.  
(See figure~(2) in the former reference
and equation~(12) in the
latter reference.)
Both of these papers 
used an incorrect
expression for the dependence of the optical depth
on the cross-section.  In equation~(5) of \cite{feh93},
and in equation~(4) of \cite{wl95}, there should be
an extra factor of $1-\cos\theta$ multiplying
the cross-section, where $\theta$ is the angle between 
annihilating photons.  This extra factor is in addition to the
dependence of the cross-section on $\theta$.  
It accounts for the fact that, if two photons are traveling
nearly parallel to each other, then they are unlikely to 
interact---in the limit that they are exactly 
parallel ($\theta=0$), they will never 
interact.  Since, approximately, $1-\cos\theta=\gamma^{-2}$,
this accounts for the difference between these authors' results
and ours.

\cite{bh97} 
presented an expression for the optical depth which is
proportional to $\gamma^{-2\alpha}/R_0$, where $R_0$ is the radius of
the burst (their equation~(41), for example).  
This is in agreement with our expressions. 
However, they then claimed that
$R_0$ is proportional to ``one or two powers'' of $\gamma$. 
They finally concluded, in their section~3.3, that 
$R_0\simeq\gamma c \delta T$ is more appropriate,
yielding $\tau\propto \gamma^{-2\alpha-1}$, and used
that for their numerical results.
However, as long as the burst is spherically symmetric 
within the angle $1/\gamma$, the only possibility is 
$R_0\simeq\gamma^2 c \delta T$. While the numerical coefficient in
this equation is uncertain, the exponent on $\gamma$ is not.
The relevant quantity is the
number of photons per unit area at the source. 
One could also use $R_0\simeq\gamma c \delta T$, which is the 
transverse size seen by a given observer, in the denominator of 
equation~(\ref{eq:tau}), but then only the photons emerging from the 
transverse area should be used in the numerator.  The number of these
photons is correspondingly reduced by $1/\gamma^2$, and our 
equation~(\ref{eq:tau}) remains valid.
In addition, we note that the dependences on cosmological redshift
are incorrect in \cite{bh97}.  The correct dependence for the
optical depth is $\tau\propto (1+z)^{2\alpha-2}$, assuming
that it is the luminosity distance which is used.
The resulting dependence on redshift of the lower limit on 
$\gamma$ is as listed in table~(1).

Our calculation is similar in spirit to that presented in
the review article of \cite{pir99} (his equation~(10)).
However, when calculating the number of photons, he used
the total number of photons emitted throughout the burst, whereas
it is more appropriate to consider only the number of
photons emitted during the variability timescale $\delta T$.
After correcting for this, his estimate of the optical depth
is similar to ours 
(assuming $\alpha$ in his equation~(10) is our $\alpha-1$, rather
than as defined in his equation~(2)).

Finally,  \cite{kp91} calculated the optical depth,
but in a scenario which is different than ours.
Specifically, they assumed that the
emitting material consists of rigid blobs which
move with a bulk Lorentz factor $\gamma$, such as 
would be the case for a relativistically
moving star.  Therefore, their limit depends only on
the luminosity and not on the timescale of variability.
They obtained 
$\tau\propto\gamma^{-2\alpha-1}$,
which is correct for their scenario, but which differs
from our result by one power of $\gamma$.

While the aforementioned papers presented more detailed
scenarios for the expansion of the burst 
than presented here, differing scenarios would change
our parameter $\hat{\tau}$ by factors which are of order
unity.  However, the dependence of the minimum $\gamma$ on 
$\hat{\tau}$ is very weak---$\gamma$ is proportional 
to  $\hat{\tau}^{1/6}$ when $\alpha=2$---and thus these factors of order
unity have little effect on the limit on $\gamma$.  
We prefer to leave the expansion scenario vague, largely
because there are too many alternatives, each of which
would produce a different factor of order unity in $\hat{\tau}$,
and none of which would significantly affect the minimum 
$\gamma$.  In addition, when we compared our result
with the more detailed model of  \cite{bh97} 
(but with their $\gamma$-dependence corrected),
we found that our lower limit on $\gamma$ exceeded theirs
by less than around 30\% for typical values of $\alpha$.

\subsection{Discussion of Two Assumptions}
\label{sec:dota}

In the above calculation of the limit on $\gamma$, 
two implicit assumptions have been made.  First, it was assumed that 
$e_{\rm max,an}>e_{\rm min}$, i.e. that the photon which annihilates
the $e_{\rm max}$ photon does not have an energy which is smaller
than the low-energy break in the photon spectrum.
In terms of observed quantities, this condition can
be written as $e_{\rm max}<m_ec^2\hat{\tau}^{1/2}$ 
(assuming that $e_{\rm min}<m_ec^2$).  
It is easily satisfied for all of the 
bursts which we have considered 
(except for the burst from which TeV photons may
have been detected;  see section~(\ref{sec:obsres}),
below).
If it is not satisfied, then the limit on $\gamma$
depends on the behaviour of the photon spectrum
below $e_{\rm min}$.  In particular, if the number
of photons is dominated by the high end of this 
part of the spectrum
(i.e. by photons whose energies
are around $e_{\rm min}$), then the condition to be optically
thin is $e_{\rm thick}<e_{\rm min}$; this leads to
$\gamma> \hat{\tau}^{1\over 4}/
(e_{\rm min}/m_e c^2)^{\alpha-1\over 4}$.
Conversely, if the number of photons is dominated by
the low end of this part of the spectrum, then the 
limit which has been calculated in this section 
in equation~(\ref{eq:gamma1}) is applicable
as long as it is this part of the spectrum which is used
to calculate $\hat{\tau}$;  i.e. in equation~(\ref{eq:tauhat}),
the quantities $f$ and $\alpha$ should be defined 
by requiring that the spectrum below $e_{\rm min}$
is $fe^{-\alpha}$.

Second, it was assumed that
the photon with energy $e_{\rm max}$ can annihilate
a second photon whose energy is equal to its own
(i.e. $e_{\rm max}>m_ec^2\gamma$), and thus
that photons which annihilate the $e_{\rm max}$ photon have
a minimum energy which is less that $e_{\rm max}$.  If,
conversely, the lower limit on $\gamma$ obtained in
equation (\ref{eq:gamma1}) does not satisfy this
constraint (i.e. if $\gamma>e_{\rm max}/m_ec^2$), then
the $e_{\rm max}$ photon can only annihilate photons
with energies larger than itself.  
Since we do not observe these photons, we can only speculate
about their existence.  We consider two alternatives.

A first alternative is that the GRB does not produce any photons
whose energies would be observed to be larger than $e_{\rm max}$ 
(independent of any optical depth considerations).
This alternative implies that the limit on 
the Lorentz factor is
\begin{equation}
\gamma>e_{\rm max}/m_ec^2 \ .
\end{equation}
However, it seems unlikely that this limit  would be valid for most
bursts: since $e_{\rm max}$ is often the energy above which
the extrapolated photon flux would be too small to be observed, it would
be overly coincidental if $e_{\rm max}$ were also the
energy above which the intrinsic photon spectrum 
(i.e. before optical depth considerations)
were cut off.

A second alternative is that this intrinsic spectrum has no
cutoff at very high energies.
We consider this to be the more realistic of the two alternatives.  
In this case, the limit calculated previously in 
equation~(\ref{eq:gamma1}) would be valid.  
However, if this second alternative is
true, then there is a more stringent bound on $\gamma$ which will
be discussed in the following section.

\section{ LIMIT B: FROM COMPTON SCATTERING OFF 
PAIR-PRODUCED e$^\pm$}\label{sec:method2}

The photons which annihilate each other produce electron-positron pairs.
These pairs can, in turn, Compton scatter other photons. 
Now, we can approximate the number of electron-positron pairs by
the number
of photons which both (i)  have energy 
greater than $e_{\rm self,an}\equiv m_ec^2\gamma$, 
sufficient to ``self-annihilate'';  
and (ii) are optically thick to pair creation.
The burst will
be optically thin if the number of these pairs is smaller than
$N_{>e_{\rm thick}}$ (as defined 
in equation~\ref{eq:ethick}). 
Equivalently, the burst will be
optically thin if
$e_{\rm self,an}>e_{\rm thick}$, i.e. if
\begin{equation}
\gamma>\hat{\tau}^{1\over\alpha+3} \ ,
\label{eq:gamma2}
\end{equation}
where $\hat{\tau}$ has been defined in equation (\ref{eq:tauhat}).
For simplicity, in the above calculation we 
implicitly took the Compton cross-section
to be 11/180 $\sigma_T$, instead of the correct value
$\sigma_T$. 
When we  use this correct value, the right-hand side of 
equation~(\ref{eq:gamma2})  is increased by the
numerical factor
$(180/11)^{1\over 6+2\alpha}$, a correction which
we will ignore. 

In table~(1),
the dependence of $e_{\rm self,an}$ on cosmological redshift is
obtained by replacing $\gamma$ by $\gamma/(1+z)$.
Finally, we note that if limit~B did not hold, then the burst
would be optically thick to all photons, and not just to those photons
which have high energies.

\subsection{Comparison With Limit From Photon Annihilation}

Limit~A is the requirement that 
$e_{\rm thick}<e_{\rm max,an}$; limit~B is the requirement
that $e_{\rm thick}<e_{\rm self,an}$.  Thus, limit~B is 
more important than limit~A if
$e_{\rm self,an}<e_{\rm max,an}$ (or, equivalently, 
$e_{\rm max}<e_{\rm self,an}$).  This is simply the
requirement that the photon with energy $e_{\rm max}$
cannot self-annihilate.
In terms of observed quantities, this condition that
limit~B be more important than limit~A may be written as 
$e_{\rm max}<m_ec^2\hat{\tau}^{1\over \alpha+3}$. 

Finally, it should be emphasized that limit~B implicitly assumes
that the intrinsic photon spectrum (before optical depth considerations)
can be extrapolated to the energy which corresponds to $e_{\rm self,an}$.
Although $e_{\rm self,an}$ is unobservable when limit~B is 
applicable, we believe that this assumption is reasonable.

\section{ LIMIT C: FROM SCATTERING OFF ELECTRONS
ASSOCIATED WITH BARYONS}

If there are baryons in the GRB, then
a third limit on $\gamma$ may be obtained by considering
the scattering of photons by electrons which are associated with
these baryons.  Several previous papers have used this limit
(e.g. \cite{sp97b}, \cite{gps00}).
Our calculation assumes that the energy in photons is less than 
$\gamma$ times the baryon rest mass energy, 
which is valid in scenarios where the baryons are cold (such
as internal shock models).
However, this assumption is not valid in some scenarios,
such as  external shock models where the electrons are heated to
equipartition with the baryons---in which case the energy in
photons is $\gamma^2$ times the baryon rest mass energy.
In addition, for these models, one must use the Klein-Nishina
cross-section instead of the Thomson cross-section which we 
use below.  Thus, the validity of this limit is restricted. 

The optical depth due to Compton scattering off of the electrons which
are associated with baryons is, with similar assumptions
which led to equation (\ref{eq:tau}),
\begin{displaymath}
\tau={\sigma_T N_{\rm baryons}
\over 4\pi(\gamma^2c\delta T)^2} \ ,
\end{displaymath}
where $N_{\rm baryons}$ is the number of baryons.  Now,
a lower limit on $N_{\rm baryons}$ can be obtained
when the total energy in photons is less than the kinetic
energy of the baryons,  i.e.
$e_{\rm min}N_{>e_{\rm min}}<N_{\rm baryons}m_pc^2\gamma$, 
where $m_p$ is the proton mass.  
With equation (\ref{eq:ngte}) for $N_{>e}$, the 
optical depth due to the minimum number of baryons is 
$\tau=\hat{\tau}\gamma^{-5}(e_{\rm min} / m_p c^2)$. 

To compare with limit~B (equation~\ref{eq:gamma2}), 
we write the resulting limit 
on $\gamma$ as follows: 
$\gamma>\hat{\tau}^{1\over\alpha+3}
[(\hat{\tau}^{1\over\alpha+3})^{\alpha-2}
(e_{\rm min}/m_p c^2)]^{1\over 5}$.
However, since $(e_{\rm min} / m_p c^2)$ is typically very
small (around $10^{-3}$ or $10^{-4}$), the quantity in
square brackets is typically smaller than unity, and
limit~C can be ignored.

\section{OBSERVATIONAL RESULTS}
\label{sec:obsres}

Table~(2) lists the relevant limits 
for a number of bursts.  
The quantity $\hat{\tau}$ which is in the table
has been defined in equation (\ref{eq:tauhat}).
Numerically,
\begin{equation}
\hat{\tau}=
2.1\times 10^{11} \cdot 
\Big[
{
(d/{\rm 7 \ Gpc})^2 (0.511)^{-\alpha+1}f_1 \over
(\delta T/{\rm 0.1 sec})(\alpha-1) 
} 
\Big] \ ,
\label{eq:numtauhat}
\end{equation}
where $f_1$ is 
the observed number of photons per second per cm$^2$ per MeV
at the energy of 1 MeV,
i.e. $f_1\equiv f$  MeV$^{-\alpha+1}$ sec cm$^2$.

Both limit~A and limit~B are listed in the table. Clearly,
only the larger limit is relevant; it is listed in boldface.
We re-emphasize that it is being assumed that the
intrinsic photon spectrum (i.e. before optical depth effects)
can be extrapolated to very high energies.  In particular,
when limit~B is relevant, the photon spectrum is extrapolated
to $e_{\rm self, an}$. However, when limit~A is relevant, the
spectrum need not be extrapolated past the observed 
energy $e_{\rm max}$.

From the first set of seven bursts in table~(2),
very high energy photons were observed. 
The parameters for the first six of these bursts were tabulated in 
\cite{bh97}.  The data originally come
from EGRET, except for GRB~910601, where COMPTEL data
give stronger constraints.
From bursts with measured redshifts, 
the redshift $z=1$ is a plausible estimate.  This corresponds
to a luminosity distance of $d=6.6$ Gpc (when $\Omega_m=0.3$,
$\Omega_\Lambda=0.7$, and $H_0=70$ km sec$^{-1}$ Mpc$^{-1}$, which
are the generally accepted values for the cosmological parameters).
Note that the luminosity distance
is nearly proportional to $(1+z)^2$.  This implies that both 
limit A and limit B are roughly proportional to $1+z$; and, for example,
the limits would be doubled if the bursts had $z=3$.
We set the variability time $\delta T=0.1$ sec,
because BATSE detects variability down to this timescale, which
is comparable to BATSE's resolution time.
The seventh burst, GRB~990123, had its redshift measured to be
$z=1.6$, which corresponds to a luminosity distance of 11.8
Gpc. 

For the remainder of the bursts in table~(2),
not many high energy photons were observed.
To estimate the quantity $f_1$ in equation
(\ref{eq:numtauhat}), we use BATSE data.  In particular,
defining ${\cal F}$ as the fluence in photons with energy
greater than 300 keV, and $T$ as the duration of the
burst during which 50\% of the fluence is observed,
we use the approximation
\begin{displaymath}
f_1={{\cal F}/300{\rm keV}\over T}\Big(
{300 {\rm keV}\over 1 {\rm MeV}}\Big)^{\alpha-1} \ .
\end{displaymath}

The middle set of bursts in table~(2) are
three GRB's which had measured redshifts, and which
had relatively large total energies.  In addition to 
the expression for $f_1$ above, we used $\delta T=0.1$ sec
and $e_{\rm max}=m_ec^2$.  With this value of $e_{\rm max}$,
it is possible that these bursts are non-relativistic, if one
assumes that the intrinsic photon spectrum is cut off above
$e_{\rm max}$.
Nonetheless, under the reasonable assumption that the intrinsic
photon spectrum continues to very high energies, then
the resulting limits are as listed in table~(2) for two 
possible values of $\alpha$.
Note that limit~B is always more significant than limit~A
for these bursts, mainly due to the low assumed value of $e_{\rm max}$.

The final two GRB's have unusual properties.
The first of these is GRB~980425,
which was a nearby low-energy burst; it was nearly coincident
with the supernova SN~1998bw \citep{paa+00}.  
We evaluate $\hat{\tau}$
for this burst using the same method as was used in
the previous three bursts with redshifts.  However, we
use $\delta T=5$ sec because the light curve for this
burst was smooth on timescales smaller than this.

The second of the unusual bursts is GRB~970417a, from which 
TeV photons may have been detected \citep{abb+00}.  
We use BATSE data, together with
$\delta T=0.1$ sec and $z=0.3$.
This value for $z$ is an upper limit based on the
opacity due to starlight \citep{abb+00}.
Then, for this burst, $e_{\rm max,an}<e_{\rm min}$,
where $e_{\rm min}\simeq m_ec^2$.  Since, in addition,
the photon spectrum is cut off below
$e_{\rm min}$, the  limit on $\gamma$ is obtained by
considering all of the photons in the burst.  
The requirement on the Lorentz factor
is thus $\gamma>\hat{\tau}^{1/4}$ (section~\ref{sec:dota}).
The resulting limit
on $\gamma$ is listed in table~(2).  
In our calculation for this burst, we assume that the 
optical depth is below unity for all photon energies up
to TeV.
\section{SUMMARY}

We derived three limits on the Lorentz factor $\gamma$,
based on the requirement that GRB's be optically
thin. Our order of magnitude calculations are unaffected
by the details of the scenario and our lower limits apply
to internal as well as external shocks, assuming spherical
symmetry within an opening angle of $1/\gamma$.
Limit~A was obtained by considering the annihilation
of pairs of photons; limit~B was obtained by considering 
the scattering of photons by pair-created e$^\pm$; limit~C
was obtained by considering the scattering of photons by
electrons which are associated with baryons.  It was shown
that, as long as the intrinsic photon spectrum (i.e. before optical
depth effects) can be extrapolated to very high energies, limit~C
is nearly always less important than limit~B.
Table~(1) 
summarizes the results for limits~A and~B.

We evaluated limits~A and~B 
for a few selected bursts
(see table~2).  Our numerical results for 
limit~A are
not very different from the results of 
\cite{wl95} and \cite{bh97}.
While correcting their dependence of optical depth
on Lorentz factor reduced the lower limit on $\gamma$ by 
a factor of a few, our use of larger distances (because
of recent redshift measurements) partially compensated
for this.  

With the anticipated launch of
the Gamma Ray Large Area Space Telescope (GLAST) in 2005, the limits
presented in this paper may be improved.  
Since cut-offs in the photon spectrum at high energies have usually not
been observed, the values of $e_{\rm max}$ which we have used in this paper
are lower limits, set by instrumental sensitivity.
GLAST will be able to detect
photons up to 300 GeV, with sensitivity much better than that of
EGRET (http://glast.stanford.edu/mission.html).  
If GLAST does not observe a cut-off, then 
compared with our assumed value for the maximum observed
energy photon from bursts with measured resdshifts (i.e. $e_{\rm max}=m_ec^2$), 
$e_{\rm max}$ would be increased by a factor of 
nearly $10^6$ .  Since our limit A is proportional to $e_{\rm max}^{1/6}$
(when $\alpha=2$), our limits could
be increased by a factor of around 10, yielding lower limits on 
Lorentz factors of nearly 1000.

\acknowledgments

We thank E. Fenimore, P. Goldreich, A. Loeb, and C. Matzner
for helpful discussions.
RS acknowledges support from the Sherman Fairchild Foundation.


\begin{center}
\begin{table}[ht!] 
\begin{center}
\caption{\textbf{Summary of main theoretical results$^1$}}
\begin{tabular}{rll}\hline
$e_{\rm max}$:&observed&maximum observed energy \\ \hline
$e_{\rm max,an}$:
&&
           minimum energy of photon which  
           annihilates \\
&
\raisebox{1.5ex}[0pt]{${(\gamma m_e c^2)^2\over e_{\rm max}}(1+z)^{-2}$}
& $e_{\rm max}$  photon \\ \hline
$e_{\rm self,an}$ :&$m_ec^2\gamma(1+z)^{-1}$&minimum energy of photon
           which  annihilates \\&& itself \\ \hline
$e_{\rm thick}$:&
           $m_ec^2(\hat{\tau}\gamma^{-4})^{1\over \alpha-1}$
           &above this energy the number  of photons \\&&
           is such 
           that, if they were assigned a Thomson \\&& 
           cross-section, they would be optically thick \\ \hline  \\
\end{tabular}
\begin{tabular}{lll}\hline
&Requirement that $\tau<1$&lower limit on $\gamma$ \\ \hline\hline
Limit A&$e_{\rm max}$ photon can escape without
	    &\\&
             annihilating other photons
	    $\Rightarrow e_{\rm thick}<e_{\rm max,an}$
	   &
            \raisebox{1.8ex}[0pt]{
           $\hat{\tau}^{1\over 2\alpha+2}(
           {e_{\rm max}\over m_ec^2}) ^{\alpha-1\over 2\alpha+2}
           (1+z)^{\alpha-1\over\alpha+1}$
                                 }
            \\ \hline
Limit B&e$^{\pm}$ pairs produced by photon annihilation 
        &\\&
          are optically thin 
           $\Rightarrow e_{\rm thick}<e_{\rm self,an}$
           &
            \raisebox{1.8ex}[0pt]{
           $\hat{\tau}^{1\over\alpha+3}
            (1+z)^{\alpha-1\over\alpha+3}$ 
                                 }
             \\ \hline
\end{tabular}
\end{center}
\begin{center}
$^1$ The quantity $\hat{\tau}$ is defined in equation~(\ref{eq:tauhat}), 
where $d$ is the luminosity distance.
\end{center}
\end{table}
\end{center}

\begin{center}
\begin{table}[ht!] 
\begin{center}
\caption{\textbf{Limits on selected bursts}}
\begin{tabular}{llllllcccc}\hline
&GRB&$f_1$&$\alpha$&${e_{\rm max}\over m_ec^2}$&$z$&$\hat{\tau}$&
Limit A: $\gamma_{\rm min}=$&
Limit B:&Ref.$^1$
\\ &&&&&&&
           $\hat{\tau}^{1\over 2\alpha+2}(
           {e_{\rm max}\over m_ec^2}) ^{\alpha-1\over 2\alpha+2} $
&   
           $\gamma_{\rm min}=\hat{\tau}^{1\over\alpha+3}$ 
&\\&&&&&&&
           $\times (1+z)^{\alpha-1\over\alpha+1}$&
           $\times (1+z)^{\alpha-1\over\alpha+3}$&\\ \hline
Bursts&910503&8.71&2.2&333&1&$3.0\times 10^{12}$&\bf{340}&300&1\\
with&910601&0.5 &2.8&9.8&1&$1.8\times 10^{11}$&72&\bf{110}&2\\
very&910814&13.5&2.8&117&1&$4.7\times 10^{12}$&\bf{200}&190&3\\
high&930131&1.95&2.0&1957&1&$7.0\times 10^{11}$&\bf{420}&270&4\\
energy&940217&.36&2.5&6614&1&$1.2\times 10^{11}$&\bf{340}&120&5\\
photons&950425&1.62&1.93&235&1&$6.0\times 10^{11}$&\bf{300}&280&6\\ 
&990123&1.1&2.71&37&1.6&$1.2\times 10^{12}$&150&\bf{180}&7\\
\hline
      &971214&.35&2&1&3.42&$2.6\times 10^{12}$&192&\bf{410}&8\\
Bursts&``\ \ \ \ ''&.1&3&1&3.42&$7.5\times 10^{11}$&64&\bf{160}&8\\
  with&980703&.08&2&1&.966&$2.7\times 10^{10}$&69&\bf{140}&8\\
 $z$'s&``\ \ \ \ ''&.02&3&1&.966&$8.0\times 10^{9}$&24&\bf{56}&8\\
      &990510&.1&2&1&1.62&$1.2\times 10^{11}$&98&\bf{200}&8\\ 
      &``\ \ \ \ ''&.03&3&1&1.62&$3.7\times 10^{10}$&34&\bf{79}&8\\ \hline
Unusual&980425&.04&2&1&.0085&$1.0\times 10^4$&4.6&\bf{6.4}&8 \\
bursts&``\ \ \ \ ''&.01&3&1&.0085&$2.9\times 10^3$&2.8&\bf{3.8}&8 \\ 
&970417a&$~-$&$-$&$2\times 10^6$&.3&$8.7\times 10^{8}$&\multicolumn{2}{c}{\bf{~~~~170}}&8,9 \\ \hline
\end{tabular}
\end{center}
$^1$References: 
(1) \cite{sbf+92}; (2) \cite{hbc+94}; (3) \cite{kwo93};
 (4) \cite{sbd+94}; (5) \cite{hur94}; (6) \cite{cds+96};
(7) \cite{bbk+99}; (8) http://cossc.gsfc.nasa.gov/batse/ ;
(9) \cite{abb+00} 
\end{table}
\end{center}

\end{document}